\documentclass[twoside]{article}
\usepackage{jwe,epsfig}

\begin{document}
\setlength{\textheight}{8.0truein}    

\runninghead{Title  $\ldots$}
            {Author(s) $\ldots$}

\normalsize\textlineskip
\thispagestyle{empty}
\setcounter{page}{1}

\copyrightheading{0}{0}{2003}{000--000}

\vspace*{0.88truein}

\alphfootnote

\fpage{1}

\centerline{\bf
RUN-TIME APPLICATION MIGRATION }
\vspace*{0.035truein}
\centerline{\bf USING CHECKPOINT/RESTORE IN USERSPACE}
\vspace*{0.37truein}
\centerline{\footnotesize
Aleksandar Tošić}
\vspace*{0.015truein}
\centerline{\footnotesize\it University of Primorska Faculty of Mathematics, Natural Sciences and Information Technologies, Koper}
\centerline{\footnotesize\it InnoRenew CoE, Livade 6a, Izola}
\baselineskip=10pt
\centerline{\footnotesize\it Koper, Slovenia}
\baselineskip=10pt
\centerline{\footnotesize\it 
aleksandar.tosic@upr.si}

\vspace*{0.225truein}
\publisher{(20.7.2023)}{(revised date)}

\vspace*{0.21truein}

\abstracts{
This paper presents an empirical study on the feasibility of using Checkpoint/Restore In Userspace (CRIU) for run-time application migration between hosts, with a particular focus on edge computing and cloud infrastructures. The paper provides experimental support for CRIU in Docker and offers insights into the impact of application memory usage on checkpoint size, time, and resources. Through a series of tests, we find that the time to checkpoint is linearly proportional to the size of the memory allocation of the container, while the restore is less so. Our findings contribute to the understanding of CRIU's performance and its potential use in edge computing scenarios. To obtain accurate and meaningful findings, we monitored system telemetry while using CRIU to observe its impact on the host machine's CPU and RAM. Although our results may not be groundbreaking, they offer a good overview and a technical report on the feasibility of using CRIU on edge devices. This study's findings and experimental support for CRIU in Docker could serve as a useful reference for future research on performance optimization and application migration using CRIU.
}{}{}

\vspace*{10pt}

\vspace*{3pt}
\communicate{to be filled by the Editorial}

\vspace*{1pt}\textlineskip    
\section{Introduction}        

In recent years, there has been a significant shift in the way computing is done. Instead of relying solely on centralized cloud-based systems, there has been a move towards decentralized edge computing systems. Edge computing systems bring computing resources closer to the end-users, reducing latency, improving performance, and enabling new use cases that were previously unfeasible. However, with edge computing comes a new set of challenges, such as resource constraints, high variability, and unreliable connectivity.

Checkpoint/Restore In Userspace (CRIU) is a powerful tool that allows the migration of running applications from one host to another while preserving their states. This tool has significant potential for use in edge computing systems, where migration of running applications may be necessary due to resource constraints or other factors. The ability to migrate running applications while preserving their state can significantly improve the reliability, availability, and fault-tolerance of edge computing systems. A few potential use case pertaining to the aforementioned use case are:

\begin{enumerate}
    \item Fault Tolerance and Disaster Recovery: CRIU can be used to provide fault tolerance and disaster recovery capabilities in edge computing systems. By migrating running applications to a different host in the event of a failure or disaster, CRIU can help ensure that critical services remain available and reduce the impact of downtime.

    \item Mobility: CRIU can be used to provide mobility capabilities in edge computing systems. By allowing running applications to be migrated between different hosts, CRIU can enable new use cases such as mobile edge computing, where services follow users as they move around.
    
    \item Dynamic Scaling: CRIU can be used to enable dynamic scaling of services in edge computing systems. By migrating running applications between hosts based on changes in demand, CRIU can help ensure that the system remains responsive and that resources are used efficiently.
    
    \item Energy Efficiency: CRIU can be used to enable energy-efficient computing in edge computing systems. By migrating running applications to hosts that are currently idle, CRIU can help reduce overall energy consumption and improve the sustainability of the system.
\end{enumerate}

One of the main points of interests is the advent of decentralized protocols for autonomous container management and orchestration. Protocols such as Caravela \cite{pires2021distributed}, and Nion Network \cite{tovsic2022blockchain} proposed a decentralized network of nodes that perform container migrations in an effort to balance resource utilization across the entire network. Nion network secures the decision making process through an efficient and scalable blockchain consensus mechanism. Unlike typical blockchain protocols, the blocks are viewed as states descriptions of all the containers under management. Adding new blocks to the chain can be considered as the state transition, while the state transition function is nodes performing planed migrations of containerized applications. 

Clearly, such networks are subject to Byzantine behaviour and while consensus mechanisms blockchains typically use tolerate such faults, the applications do not. These faults are traditionally addressed by introducing redundancy where possible and/or backup. In this use-case, redundancy would require more nodes to run the containerized application and serve as ready backups. Evidently, this does raise concern over the efficiency of the entire network due to additional compute resource utilization. On the other hand, backups only impose additional storage requirements but have to be frequent to avoid loss loosing states.

CRIU can be used to regularly snapshot applications and commit these snapshots to other nodes. In an event of a fault on the host machine(node), the protocol can detect it, and restore the application on another node. However, given the experimental nature of CRIU, the exact resource requirements for creating checkpoints and restoring them needs to be well understood. Moreover, studying the telemetry while performing such operations can give a better understanding into the limitations of run-time migrations.

In this paper, we explore the potential benefits of using CRIU in edge computing systems using docker experimental features that support CRIU \cite{merkel2014docker,10.1145/3357526.3357542}. We present an empirical study of the performance impact of CRIU on Docker containers and investigate the feasibility of using CRIU in edge devices. Specifically, we examine the impact of the size of the memory allocation of the host system on the time to checkpoint and restore, as well as the resources consumed during the process.

Our results demonstrate the potential benefits of using CRIU in edge computing systems, providing insights into the performance impact of using this tool in these settings. We believe that this research will contribute to the growing body of work exploring the use of checkpointing and migration in edge computing systems and will pave the way for further research into the practical applications of CRIU in this domain.

\section{Related work}
CRIU, which stands for "Checkpoint/Restore In Userspace," is an open-source software tool for performing live migrations of Linux processes. It was first introduced in 2011 by Pavel Emelyanov as a way to checkpoint and restore individual processes in Linux user space, with the goal of enabling faster live migrations for cloud computing workloads. Since then, CRIU has gained popularity in both cloud and edge computing settings, where it allows for efficient and flexible management of distributed applications.

Since it's inception the project has matured considerably. One of the earliest studies analyzing the performance of checkpoint and restore was published by Chen Yang \cite{chen2015checkpoint}. Yang analyzed the time it took to checkpoint and restore a process and unidentified process memory allocation to be the most impactful factor. However, Yang's study is very limiting as only small processes were used, and no other telemetry such as CPU utilization were analyzed while performing migrations.

Oh, SeungYong, and JongWon Kim experimented the use of CRIU for live migrations of containerized applications \cite{oh2018stateful}. In their study, they analyze the performance of CRIU with a focus on latency analysis in transferring checkpoints between hosts and were less concerned about the resources and time needed to perform these tasks.

A similar study in which CRIU was used to improve the availability of docker swarm, a cluster orchestration solution provided by Docke \cite{huang2017enhancing}. Their findings report that higher availability can be achieved by periodically checkpointing the states of containers within the swarm and restoring them to provide higher availability. Similarly to the study of Oh, SeungYong, and JongWon Kim, their analysis is focused on the memory use and storage on the network file system (NFS), and did not study the impact on the computing resource utilization while performing CRIU. Moreover, their study is very limiting with very few scenarios and container configurations analyzed. 

A more recent study by Adityas et. al. analyzed the resource utilization of the host machine while performing checkpoint and restore \cite{widjajarto2021live}. In their study, they report the expected memory allocation of the container to have the most impact on the time it takes to perform checkpoint and restore. Interestingly, their study also reports on the CPU consumption while performing these tasks. However, instead of extensively testing various containers, the authors chose a scenario based approach. They identified four three main, namely \emph{One way migration between two hosts one direction at a time}, \emph{two way migrations using one service}, and \emph{two way migrations using three services}. While the preliminary results reported do offer some insight, the study is very limited to the aforementioned scenarios. The study would benefit a more detailed analysis of multiple services with various memory allocations and a much larger sample size.

Previous research suggests there is the use of CRIU is ubiquitous in edge and micro-service architectures. However, the lack of an in-depth performance analysis that highlights the limitations is evident. Our study aims to address the gap and pave the way for future use of CRIU for run-time application migrations on the edge.

\section{Experimental Results}

When CRIU is used to checkpoint a running application, it performs the following steps:
\begin{enumerate}

   \item Image creation: CRIU creates an image of the process that needs to be checkpointed. This image contains information about the process's memory, open files, network connections, and other relevant information. The image is created in a checkpoint file.

   \item Freezing: Once the image is created, CRIU freezes the process. Freezing is the process of pausing the execution of the process so that its state does not change while the checkpoint is being created.

   \item Image dumping: CRIU dumps the image of the process into the checkpoint file. This process involves saving the state of the process's memory, CPU registers, file descriptors, network connections, and other relevant information.

   \item  Thawing: After the image has been dumped, CRIU unfreezes the process and allows it to resume execution.
 
\end{enumerate}

The checkpoint file created by CRIU contains all the information necessary to restore the process's state at a later time. When the process needs to be restored, CRIU reads the checkpoint file and recreates the process from the saved state. This process involves creating a new process, restoring the process's memory, CPU registers, file descriptors, network connections, and other relevant information, and then resuming the process's execution.

The ability to store and restore the state of a running application is possible due to the design of the Linux kernel. Linux uses a virtual memory system to manage memory, which allows processes to have their own virtual address space. This means that each process has its own view of memory, and the kernel can manage the physical memory separately for each process. When a process is checkpointed, its virtual memory state is saved, including the contents of the process's memory and its CPU registers. This allows the process to be restored to the same state later, even if it is running on a different host or at a different time.

\begin{figure}[ht!]
\centering
\includegraphics[width=1\textwidth]{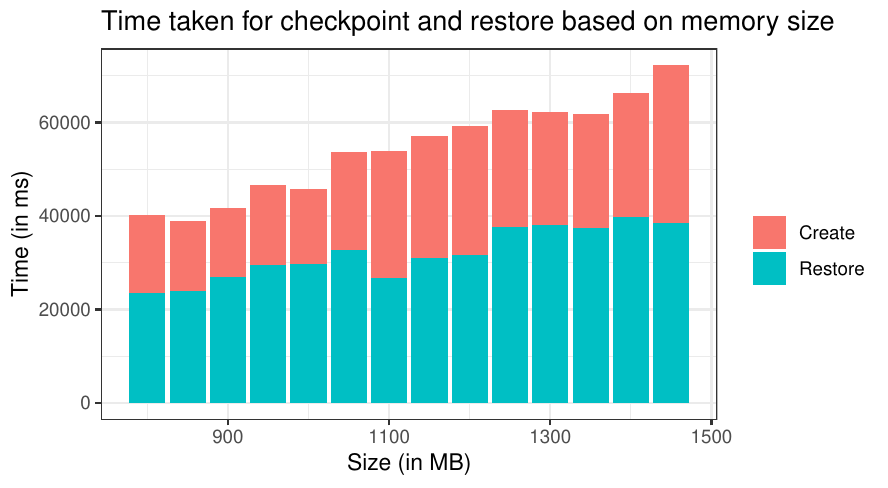}
\fcaption{CRIU checkpoint and restore timings relative to the amount of memory used by the docker container.}
\label{fig_timings}
\end{figure}

Memory is obviously the main factor that impacts the performance of CRIU.
To get a better understanding relationship between the size of virtual memory applications require and CRIU performance, a program written in Rust, was deployed as a docker container, the program allocates a given amount of memory and runs using the Ubuntu 20.04 as the base image. After the initial allocation, an external process checkpoints the container and restores it after while measuring the timings for individual tasks. Figure \ref{fig_timings} illustrates the timings (checkpoint and restore) observed for various memory allocation sizes.

We can observe that the time it takes to perform a checkpoint is proportional to the amount of memory the container had allocated. While this is not surprising, it does set some boundaries on how checkpointing can be used.

Moreover, the analysis shows that CPU utilization does not seem to scale with the size of virtual memory allocated during the checkpoint and restore phases. This suggests that CPU utilization may not be a limiting factor in checkpointing applications, and the trade-off between checkpointing and performance may be acceptable for stateless applications with limited resources available on edge devices.

Note that the container has to be paused in order to dump the pages of virtual memory and maintain consistency. Checkpoints of applications, which require high availability and constant up-time(i.e. servers) do not seem feasible. However, micro-services and stateless applications seem reasonable. The absence of global state in applications greatly reduces the memory used making CRIU a viable approach for scalability, and fault tolerance.

However, edge computing solutions have to consider limited resources available on edge devices. While the time it takes to perform a checkpoint and restore may be a trade-off worth taking for stateless applications, the CPU utilization on the host device may be a limiting factor.

\begin{figure}[ht!]
\centering
\includegraphics[width=1\textwidth]{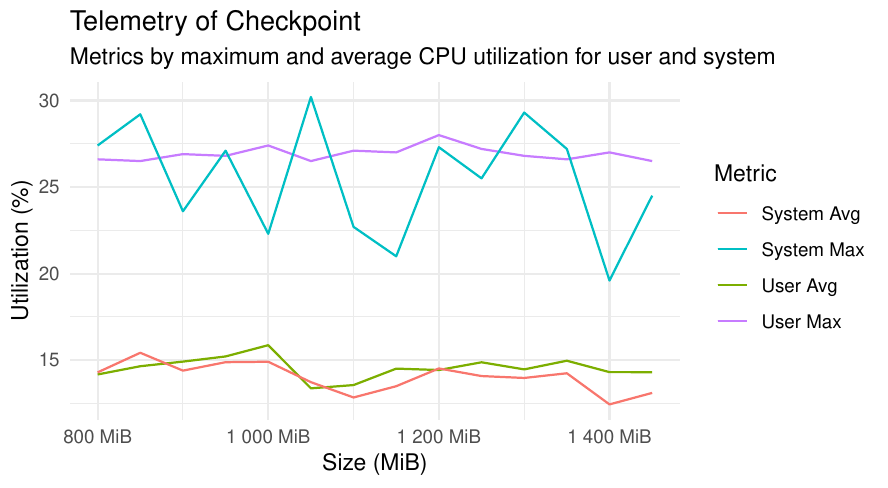}
\fcaption{CRIU checkpoint and restore timings relative to the amount of memory used by the docker container.}
\label{fig_checkpoint}
\end{figure} 

\begin{figure}[ht!]
\centering
\includegraphics[width=1\textwidth]{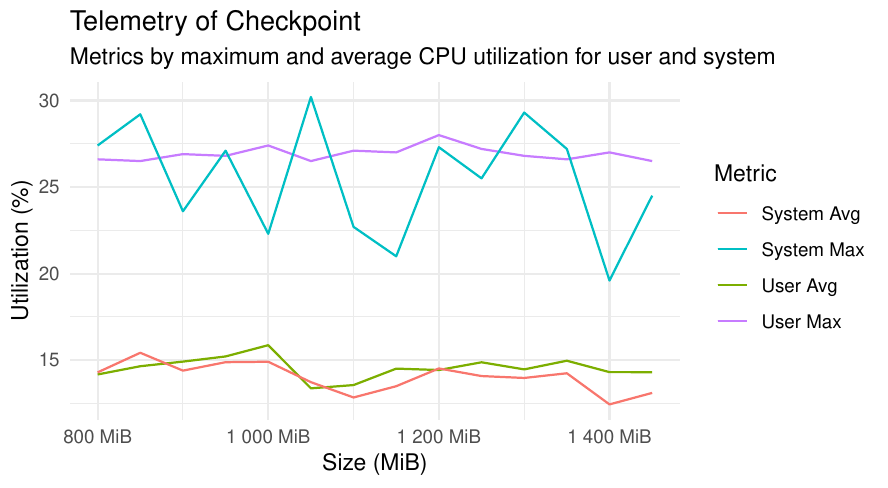}
\fcaption{CRIU checkpoint and restore timings relative to the amount of memory used by the docker container.}
\label{fig_restore}
\end{figure}

To gain some insight into the CPU utilization of the host dockerd service and CRIU we monitor system telemetry when performing checkpoint and restore tasks. Figure \ref{fig_checkpoint} illustrates the CPU utilization in \% both of the user running the checkpoint as well as the overall system. We can observe that while the memory allocation of the application directly impact the time to perform these tasks, the CPU utilization does not seem to scale with the size of the virtual memory allocated. Moreover, the same can be concluded for the restore phase observed in Figure \ref{fig_restore}.

Overall, the analysis highlights the potential benefits and limitations of using CRIU for checkpointing applications and provides insights into the trade-offs between performance, memory usage, and fault tolerance.

\section{Conclusion}
In conclusion, we have investigated the capabilities and limitations of the CRIU tool for checkpointing and restoring containers in the context of edge computing. Our findings demonstrate that CRIU provides a robust solution for stateless applications and microservices where fault tolerance and scalability are essential. However, the size of the virtual memory allocated by the application greatly impacts the time required to perform checkpoint and restore operations. Thus, edge computing solutions should consider the trade-off between checkpointing frequency and application performance. Additionally, while the CPU utilization does not seem to scale with the size of the virtual memory allocated, it remains a limiting factor in the deployment of checkpointing on edge devices with limited resources.

Future work can explore techniques to reduce the memory footprint of the container, enabling more frequent checkpoints without affecting the application's performance. Furthermore, it would be interesting to investigate the use of CRIU for more complex stateful applications and the integration of CRIU with other container orchestration tools. Overall, our study contributes to the understanding of the capabilities and limitations of CRIU in the context of edge computing and provides insights for the deployment of checkpointing solutions in resource-constrained environments.

\bibliographystyle{plain}
\bibliography{ref}

\end{document}